\documentclass[12pt]{article}
\usepackage{graphicx}
\usepackage{amsfonts}

\newcommand{\keywords}{\par \vskip 8pt{\bf Keywords }}

\newcommand{\be}{\begin{eqnarray}}
\newcommand{\ee}{\end{eqnarray}}
\newcommand{\ben}{\begin{eqnarray}}
\newcommand{\een}{\end{eqnarray}}

\newcommand{\Tr}{\,{\rm Tr}\,}

\begin{document}

\title{Generalized thermostatistics and mean-field theory}
\author{Jan Naudts\\
  \small Departement Natuurkunde, Universiteit Antwerpen UIA,\\
  \small Universiteitsplein 1, 2610 Antwerpen, Belgium\\
  \small E-mail {Jan.Naudts@ua.ac.be}
}

\date{August 2003}

\maketitle

\begin{abstract}
The present paper studies a large class of temperature-dependent
probability distributions and shows that entropy and
energy can be defined in such a way that these probability
distributions are the equilibrium states of a
generalized thermostatistics, which
is obtained from the standard formalism by deformation of exponential
and logarithmic functions. Since this procedure is non-unique,
specific choices are motivated by showing that the resulting
theory is well-behaved. In particular, with the choices made in the present paper,
the equilibrium state of any system with a finite number
of degrees of freedom is, automatically, thermodynamically stable
and satisfies the variational principle.

The equilibrium probability distribution of open systems
deviates generically from the Boltzmann-Gibbs distribution.
If the interaction with the environment is not too strong then
one can expect that a slight deformation of the exponential
function, appearing in the Boltzmann-Gibbs distribution, can reproduce
the observed distribution with its temperature dependence.  An example of a system,
where this statement holds, is a single spin of the Ising chain.
However, because all systems of the present generalized thermostatistics
are automatically stable, one must not expect that all open systems can
be described in this way. Indeed, systems exhibiting a phase transition in
their thermodynamic limit, can be unstable even when the interaction
with the environment is weak. Therefore, their equilibrium probability
distribution cannot be described by a simple deformation of the Boltzmann-Gibbs
distribution. In order to be able to handle such systems as well the second part of the paper
discusses a further extension of the class of probability distributions
using mean-field techniques.

Connections are discussed that exist between the present formalism,
Tsallis' thermostatistics, and the superstatistics formalism of Beck
and Cohen. In particular, the present generalization sheds some light onto the historical
development of the Tsallis formalism. It is pointed out that
temperature dependence of the Tsallis distribution has hardly
been verified experimentally.


\end{abstract}

\section{Introduction}

Statistical physics is primarily concerned with (nearly) closed
systems. A correct description of open systems, i.e.~systems
weakly interacting with their environment, requires that the
latter is taken into account explicitly. The present paper
attempts to generalize the thermostatistical formalism to open
systems without the need for including the environment. Such a
generalization may fill the gap between the narrow context for
which the theory of statistical physics is developed and the
vast domain of its applications. Indeed, often used tools, like
temperature-dependent Hamiltonians and mean-field
approximations, are not so well founded in the standard
formalism. The main part of the paper explores how far one can go
in modeling the temperature dependence of open systems.
In doing so, Hamiltonians are {\sl not} allowed to depend on temperature.
Rather, the Boltzmann-Gibbs distribution is replaced by more general
probability distributions and the microscopic definition of
entropy is modified accordingly.
Only the last sections include temperature-dependent Hamiltonians
as a further extension of the formalism, more suited for
unstable systems.

A general formalism like the one developed here helps also to improve understanding of
more specific generalizations of the Boltzmann-Gibbs formalism. The present paper
discusses in particular the formalism of non-extensive thermostatistics,
first introduced in \cite {TC88}. By looking from an eagle's point of
view several peculiarities of this formalism get a new interpretation.
Let me mention the role of escort probabilities, historical problems with
average energy without proper normalization and with the definition of temperature,
and the occurrence of thermodynamic instabilities.
Also the relation with the recently introduced superstatistics \cite {BC03}
is discussed. In this way the
present paper contributes to clarify the relation between mainstream
statistical mechanics and its generalizations.


It is of course very important that the present generalized formalism has
physical applications. Nevertheless, no such applications will be discussed.
The literature of non-extensive thermostatistics claims a large number
of applications --- see e.g.~\cite {ST99}. It is argued below that
these applications are candidate applications of the present formalism.
However, in most, if not all, of these papers the authors verify the shape of the
equilibrium probability distribution, but not its temperature dependence.
The present paper stresses the point that in a predictive theory of
thermostatistics both aspects are equally important. Let us now explain
this point.
 

Gibbs' postulate implies that microstates with higher energy have
a lower probability, according to the formula
\be
p_k=\frac{1}{Z(T)} \exp(-H_k/T).
\label{Gibbs}
\ee
In this expression $p_k$ is the probability of the microstate labeled
$k$. It satisfies
$p_k\ge 0$ and $\sum_kp_k=1$. $H_k$ is the corresponding energy
and $T$ is the temperature
(Boltzmann's constant is taken equal to 1).

%

Mathematically, it is clear that any probability distribution 
function (pdf) with non-vanishing probabilities can be written
in the form (\ref{Gibbs}). This requires only an appropriate
definition of the notion of energy. Indeed, given $p_k$, one can 
put
\be
H_k=-T\ln(p_k).
\label{HDef}
\ee
But, the Boltzmann-Gibbs distribution (\ref{Gibbs}) contains 
more information than the mere statement that high energy states
have an exponentially small probability. It also predicts how 
probabilities $p_k$ change with temperature $T$. Once 
temperature is taken into account a
definition of energy, like (\ref {HDef}), is not acceptable 
because in general energy levels $H_k$, defined in such a way,
are temperature dependent. One concludes that the Boltzmann-Gibbs
distribution (\ref{Gibbs}) is very specific once temperature
dependence is taken into account. In the terminology of the present
paper it is a temperature-dependent probability distribution.

The main tool of the paper is the notion of $\kappa$-deformed
logarithmic and exponential functions, a definition of which is given in Section III.
The idea of considering density functions as deformed exponential functions goes back to
\cite {TC94}. The name of $\kappa$-deformed functions has been used explicitly
in \cite {KG01}.
One could of course replace
the Boltzmann-Gibbs distribution (\ref{Gibbs}) by an arbitrary
temperature-dependent probability distribution. However, chances are
small that in such a generality one could develop a complete
thermostatistical theory. Therefore it is obvious to require that
the function, which replaces the exponential function
in (\ref{Gibbs}), satisfies some additional properties.
In the first place it should be an increasing function because
states with a higher energy should be less probable. An additional
property of the exponential function is convexity (i.e.~the second
derivative is positive). Functions satisfying both properties
(plus some technical conditions) are called $\kappa$-deformed
exponentials and are denoted $\exp_\kappa(x)$. Doing so has the
advantage that well-known expressions from standard statistical physics
are recognized on sight in the generalized formalism.


The next section starts with a discussion about the relevance of
pdfs, other than that of Boltzmann-Gibbs.
In the third section a new approach, based on $\kappa$-deformed logarithmic
and exponential functions, is proposed. Section 4 discusses the
Ising chain as an example motivating
the generalized formalism following in section 5.
Further examples are given in section 6. This includes a discussion of
Tsallis' thermostatistics.
Generalized mean-field models
are treated in section 8, using the notion of probability
dependent variables introduced in section 7. Finally,
a short discussion follows in section 9. The paper ends with
two appendices, one about the Ising model, the other about
thermodynamic stability.

\section{Evidence of limited validity of the
Boltz\-mann-Gibbs distribution}


Consider an open system described by probabilities $p_{k\gamma}$,
where $k$ labels the microstates of the system, and $\gamma$ the microstates
of the environment. Assume these probabilities are of the form (\ref{Gibbs}),
i.e.~there exist energy levels $H_{k\gamma}$ such that
\ben
p_{k\gamma}=\frac{1}{Z(T)}\exp(-H_{k\gamma}/T).
\een
Write these energies in the form
\ben
H_{k\gamma}=H_k+H_\gamma^*+V_{k\gamma}.
\een
Then the reduced system without the environment is described by the probabilities
\ben
p_k
=\sum_\gamma p_{k\gamma}
=\frac{1}{Y(T)}\exp(-(H_{k}+\langle V_k\rangle^*)/T),
\een
with the non-linear Kolmogorov-Nagumo averages \cite{KA30,NM30} $\langle V_k\rangle^*$
given by
\ben
\langle V_k\rangle^*
=-T\ln\left(\sum_\gamma p^*_\gamma\exp(-V_{k\gamma}/T)\right),
\een
with
\ben
p_\gamma^*=\frac{1}{Y^*(T)}\exp(-H_{\gamma}^*/T),
\een
and with appropriate normalizations $Y(T)$ and $Y^*(T)$.
Clearly, the reduced probabilities $p_k$ are {\sl not} of
the form (\ref{Gibbs}), since the effective energies $H_{k}+\langle V_k\rangle^*$
depend on temperature. The only exception is when the interaction $V_{k\gamma}$
between system and environment vanishes.
One concludes that the Boltzmann-Gibbs distribution (\ref{Gibbs})
can only be universally correct in the limit that
interactions of the system with its environment
are negligible. Note that for systems of statistical
mechanics the interactions cannot vanish because then
the system is no longer open. Indeed, the
dynamics of closed systems predicts conservation of energy.
As a consequence, the state of the isolated system
is not described by (\ref{Gibbs}) but by a probability
distribution concentrating on a single value of the energy.

The above reasoning shows that temperature-dependent effective Hamiltonians
are rather generic. They occur in many situations and are used so frequently
in applied statistical physics that it is a labor of  Sisyphus to quote all occurrences.
Nevertheless, they do not receive much attention
in the theoretical foundation of statistical physics.


In the above discussion of open systems the assumption is made
that the totality of system plus environment is correctly described
by the Boltzmann-Gibbs distribution. This is not a severe limitation.
Indeed, a characteristic of systems of statistical physics
with many degrees of freedom is that the
number of energy levels $H_k$ in an interval $[E,E+\Delta E]$
increases exponentially with increasing energy $E$. As a consequence,
the typical equilibrium pdf $p$ has the property
that the only microstates that contribute significantly have energy levels $E_k$ approximately
equal to the average value $U=\sum_kp_kH_k$, in the sense that
the sum of $p_k$ over these states is close to 1. This is the basis
for the equivalence of ensembles --- see \cite {LPS94a,LPS94b,LP95}
for a mathematical formulation of this property. In information theory
\cite {GR77} this property is known as equipartition theorem.

The equivalence of ensembles, when valid, implies that the actual
form of the pdf is not very crucial. It is only needed
for microstates with energy level $H_k$ in the vicinity
of the average energy $U$. In particular, most
systems of statistical mechanics cannot be used to test the validity
of the Boltzmann-Gibbs distribution.

\section{A new approach}


The present study starts from a family of temperature-dependent
probabilities, such as could be observed for an open system.
If this family is
of a certain form then one can construct a thermostatistic formalism that
produces these probabilities as the temperature-dependent
equilibrium distribution.


In order to generalize (\ref{Gibbs}), the recently \cite {NJ02} introduced
notions of $\kappa$-deformed exponential and logarithmic functions are
used. A $\kappa$-deformed logarithmic function is denoted $\ln_\kappa(x)$
and is a strictly increasing concave function defined for all positive $x$,
normalized such that $\ln_\kappa(1)=0$, and with a possible divergence
at $x=0$, mild enough so that the integral
\be
F_\kappa(x)=\int_1^x{\rm d}y\,\ln_\kappa(y)
\label{fkappa}
\ee
converges for $x=0$.
This definition is not very restrictive. A simple example of a $\kappa$-deformed logarithm is
$\ln_\kappa(x)=3\sqrt{x}-3$. Examples of families of
$\kappa$-deformed logarithms follow below --- see (\ref{tsallisln}, \ref{kaniadakisln}).

The inverse function of a $\kappa$-deformed logarithm
is a $\kappa$-deformed exponential function and is denoted $\exp_\kappa(x)$.
If $x$ is not in the image of the $\kappa$-deformed logarithm then $\exp_\kappa(x)$ is
taken equal to 0 when $x$ is too small, and equal to $+\infty$ when $x$
is too large.


Fix now a $\kappa$-deformed exponential function $\exp_\kappa(x)$. Note that
throughout the paper this function does {\sl not} depend on temperature.
The obvious generalization of (\ref{Gibbs}) is now
\be
p_k(T)=\exp_\kappa(G(T)-H_k/T),
\label{nonGibbs}
\ee
where the function $G(T)$ is chosen such that normalization $\sum_kp_k=1$
is satisfied.
The alternative of putting normalization into
the partition sum $Z(T)$, i.e.
\be
p_k(T)=\frac{1}{Z(T)}\exp_\kappa(-H_k/T)
\label{wronggen}
\ee
is discussed later on. Expressions (\ref{nonGibbs}) and
(\ref{wronggen}) are of course equivalent in case the deformed
exponential is replaced by the standard exponential function. In
fact, the difference between the two expressions is only
relevant when temperature dependence is considered.
The choice (\ref{nonGibbs}) is the simpler one
because it yields immediately an expression for the quantity
$\ln_\kappa(p_k)$, needed in the context of the definition of
entropy (see below).

\section{Ising example}

\begin{figure}
\includegraphics[width=10cm]{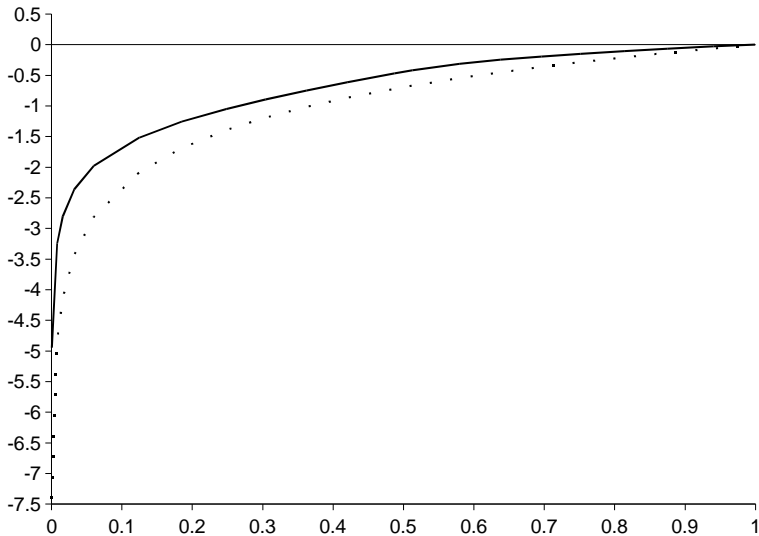}
\caption{
Deformed logarithm for the Ising model with $J/\Delta=0.25$,
according to (\ref{defloglow}, \ref{defloghigh}, \ref{uchoice}).
The dotted line is the natural logarithm.
}
\end{figure}

The following example illustrates well the point of view of the
present paper. The Ising chain in an external field can be solved exactly
in the thermodynamic limit. The result for the occupation probability
of the two levels of the spin at the origin is \cite {KW41}
\be
p_\pm(T)=\frac{1}{2}\left(1\pm\frac{y(T)}{\sqrt{1+y(T)^2}}\right),
\label{isingpdf}
\ee
with
\ben
y(T)=\sinh\left(\frac{\beta\Delta}{2}\right)e^{\beta J}.
\een
In the latter expression $\Delta>0$ is the energy level splitting due to the
external field, $J\ge 0$ is the interaction energy, and $\beta=1/T$
is the inverse temperature.


The question under study is whether there exists a thermodynamical
formalism which reproduces the temperature-dependent probabilities $p_\pm(T)$
without reference to the environment, which in this example is
formed by all other spins of the Ising chain. The obvious definition
of internal energy $U$ is
\be
U=-\frac{1}{2}[p_+(T)-p_-(T)]\Delta.
\label{isingU}
\ee
If there is no interaction with the environment (i.e. $J=0$) then
the pdfs (\ref{isingpdf}) are of the Boltzmann-Gibbs form. In this case,
Shannon's measure of information
\be
I^{\rm Shannon}(p)=-\sum_{k=\pm}p_k\ln(p_k)
\label{shent}
\ee
is an adequate expression for the entropy $S$. Indeed,
it is well-known that the Boltzmann-Gibbs distribution
is thermodynamically stable in the standard formalism
of statistical physics based on (\ref{shent}). For
further use let us verify that temperature $T$ corresponds
with the thermodynamic notion of temperature.
A short calculation using (\ref{shent}) and
the Boltzmann-Gibbs distribution gives
\ben
S=\beta U+\ln\left(\cosh\left(\frac{\beta\Delta}{2}\right)\right),
\een
so that
\ben
\frac{{\rm d}\,}{{\rm d}\beta}S
&=&U+\beta\frac{{\rm d}\,}{{\rm d}\beta}U
+\frac{\Delta}{2}\tanh\left(\frac{\beta\Delta}{2}\right)\cr
&=&\beta\frac{{\rm d}\,}{{\rm d}\beta}U.
\een
Hence, the thermodynamic relation
\be
\frac{1}{T}=\frac{{\rm d}S}{{\rm d}U}
\label{Tdef}
\ee
is fulfilled.


If the spin interacts with its environment (i.e.~$J\not=0$) then
the pdfs (\ref{isingpdf}) are not of the Boltzmann-Gibbs form,
but can be written in the form (\ref {nonGibbs}), at least when
$J/\Delta$ is not too large --- see Appendix A. For the
proof of this statement it is essential that
inverse temperature can be written as a monotonic function of the
probabilities $p_+$ and $p_-$.
Fig.~1 compares the deformed logarithm, used in this case, with the
natural logarithm. In what follows will be shown that, with
an obvious generalization of the definition of entropy (\ref{shent}),
the thermodynamic relation (\ref{Tdef}) still holds.
The definition of internal energy (\ref{isingU}) is not modified.

\section{Generalized thermostatistics}
\label{genstat}


Given a temperature-dependent family of probabilities $p_k(T)$ of the form
(\ref{nonGibbs}), a generalized
thermostatistics, which for all temperatures yields (\ref{nonGibbs}) as the
equilibrium distribution, is formulated as follows.

Let  energy as a function of temperature be defined by
\ben
U(T)=\langle H\rangle_{p(T)}\hbox{ with } \langle H\rangle_p=\sum_kp_k H_k
\label {energconstr}
\een
and let entropy as a function of temperature be defined by
\be
S(T)&=&\langle\langle I\rangle\rangle_{p(T)}
\hbox{ with }\cr
\langle\langle I\rangle\rangle_p&=&\sum_k\int_0^{p_k}{\rm d}x\,[-\ln_\kappa(x)-F_\kappa(0)].
\label{entdef}
\ee
In this expression, $F_\kappa(0)$ is a constant defined by (\ref{fkappa}).
The extended class of entropy functionals of the form (\ref{entdef}) has been studied in
\cite {NJ02,NJ03}.
A short calculation gives
\ben
\frac{{\rm d}S}{{\rm d}T}
&=&\sum_k\left[-\ln_\kappa(p_k)-F_\kappa(0)\right]\frac{{\rm d}p_k}{{\rm d}T}\cr
&=&\sum_k\left[-G(T)+H_k/T-F_\kappa(0)\right]\frac{{\rm d}p_k}{{\rm d}T}\cr
&=&\frac{1}{T}\frac{{\rm d}U}{{\rm d}T}
\een
--- to see the latter, use that
\ben
\sum_k\frac{{\rm d}p_k}{{\rm d}T}=0.
\een
This proves the thermodynamic relation (\ref{Tdef}).

Next, introduce so-called {\sl escort} probabilities \cite {BS93}, defined by
\ben
P_k(T)=Z(T)^{-1}\exp'_\kappa(G(T)-H_k/T)
\een
with $\exp'_\kappa(x)$ the derivative of $\exp_\kappa(x)$ and with
\ben
Z(T)=\sum_l\exp'_\kappa(G(T)-H_l/T)
\een
These escort probabilities have been introduced in \cite{TMP98}
for the special case of Tsallis' thermostatistics.
From $\sum_kp_k=1$ follows
\ben
\frac{{\rm d}G}{{\rm d}T}
&=&-T^{-2}\sum_kP_k(T)H_k.
\een
Using this result one calculates
\ben
\frac{{\rm d}U}{{\rm d}T}
&=&\sum_kH_k\exp'_\kappa(G(T)-H_k/T)\left(
\frac{{\rm d}G}{{\rm d}T}+T^{-2}H_k
\right)\cr
&=&Z(T)T^{-2}\sum_kP_k(T)\left[H_k-\sum_lP_l(T)H_l\right]^2\cr
&\ge&0.
\een
Hence, energy is an increasing function of temperature,
as it should be.
Together with (\ref{Tdef}), this shows that entropy $S(U)$
as a function of energy $U$ is concave. In other words,
the system is thermally stable.


There exists also a variational principle,
satisfied by the probabilities $p_k$.
Given any pdf $r$, consider the
problem of maximizing information content $\langle\langle I\rangle\rangle_r$
under the constraint that average energy $\sum_kr_kH_k$
equals a given value $U$. Introduce Lagrange multipliers
$\alpha$ and $\beta$ and minimize the expression
\ben
\beta\sum_k r_kH_k-\langle\langle I\rangle\rangle_r-\alpha\sum_kr_k.
\een
Variation w.r.t.~$r_k$ gives the condition
\ben
0=\beta H_k+\ln_\kappa(r_k)+F_\kappa(0)-\alpha.
\een
Comparison with (\ref{nonGibbs}) shows that $r=p$, $\beta=1/T$,
and $\alpha=F_\kappa(0)+G(T)$,
is a solution of this variational principle.

Assume now that $M$ is any other other observable of the system,
e.g., total magnetization of a spin system. Then thermodynamic
stability \cite {CHB85} requires not only that entropy $S$ is concave
as a function of internal energy $U$ but also as a function
of the average value $\langle M\rangle$.
That this property holds is shown in Appendix B.

\section{Examples}


The example of a single spin of the Ising chain has been 
discussed above. The pdfs (\ref{isingpdf}) are of the form 
(\ref{nonGibbs}). Therefore, the formalism developed above is
applicable. One concludes that the exact equilibrium
distribution of a weakly interacting Ising chain in the thermodynamic limit, when
restricted to a single spin variable, is an equilibrium 
distribution of generalized thermostatistics.

\subsection {Tsallis' thermostatistics}

Make the following specific choice of $\kappa$-deformed logarithm
\be
\ln_\kappa(x)=\frac{q}{q-1}\left(x^{q-1}-1\right),
\label{tsallisln}
\ee
where $q$ is a parameter, which for technical reasons should lie between 0 and 2.
The inverse function is
\ben
\exp_\kappa(x)=\left[1+\frac{q-1}{q}x\right]_+^{1/(q-1)}.
\een
The notation $[x]_+=\max\{0,x\}$ is used.
A short calculation shows that (\ref{entdef}) reduces to
\be
\langle\langle I\rangle\rangle_p=\frac{1}{q-1}\left(1-\sum_kp_k^q\right),
\label{tsalen}
\ee
which is Tsallis' $q$-entropy \cite {TC88}.
The temperature-dependent probabilities $p_k$, given by equation
(\ref{nonGibbs}), read
\be
p_k=\left[
1+\frac{q-1}{q}\left(G(T)-\frac{H_k}{T}\right)
\right]_+^{1/(q-1)}.
\label{tsaldis}
\ee


The distribution introduced by Tsallis \cite {TC88},
in its original version, reads in the  present notations
\be
p_k=\frac{1}{Z}\left[1-\beta^*(q-1) H_k\right]_+^{1/(q-1)},
\label{tsallisprob}
\ee
with $Z$ a normalization constant, and
$\beta^*$ a Lagrange multiplier related to inverse temperature.
It is of course possible to write (\ref{tsaldis}) in the form
of (\ref{tsallisprob}). The identification gives
\ben
\beta^*&=&\frac{1}{T}\frac{1}{q+(q-1)G(T)}\cr
Z&=&\left[1+\frac{q-1}{q}G(T)\right]^{1/(1-q)}.
\een
Note that $G(T)$ can be eliminated to obtain
\ben
\beta^*=\frac{1}{qT}Z^{q-1}.
\een
Clearly, the Lagrange multiplier $\beta^*$ does not have its
usual value of $1/T$, but depends in a rather complicated
way on temperature. This difficulty was not understood
in the early days of Tsallis' thermostatistics. In fact,
it took about ten years \cite {AMPP01} before the question
of defining temperature was settled.


It is interesting to sketch briefly the evolution that the
formalism has undergone. Soon after the original paper 
\cite {TC88} appeared, it was proposed \cite {CT91} to replace the constraint
$\sum_kr_kH_k=U$, used in the variational principle,
with a constraint of the form
\be
\sum_kr_k^qH_k=U.
\label{nonlinenerg}
\ee
The effect of this change is a pdf
of the form (\ref {tsallisprob}), with $q-1$ replaced by $1-q$.
The main advantage of the above constraint is that the usual
relation $\beta=1/T$ between Lagrange multiplier and thermodynamic
temperature $T$, defined by (\ref{Tdef}) is restored,
however, at the cost of changing the definition of thermodynamic
energy $U$. Both papers \cite {TC88,CT91} have been criticized \cite {RJD95,TC95},
the first because of the apparent lack of thermodynamic stability,
the latter because internal energy $U$, as given by (\ref{nonlinenerg}),
is not the average of the energy levels $H_k$.
Later attempts to cure the formalism \cite {TMP98,AMPP01}
(introducing escort probabilities and changing the definition
of temperature) solved the problems only partly.
For a recent discussion, see \cite {WT02}.

\subsection {Superstatistics}

In a recent proposal \cite {WW00,BC01,BC03,Be03} the Boltzmann-Gibbs formalism is generalized
by replacing the Boltzmann factor $\exp(-\beta E)$ by an average over a
range of inverse temperatures $\beta$
\be
B(E)=\int_0^\infty{\rm d}\beta\,f(\beta)e^{-\beta E}.
\label {superstat}
\ee
The resulting formalism is called {\sl superstatistics}.
With an appropriate choice of distribution $f(\beta)$ the Tsallis
distribution (\ref {tsallisprob}) is recovered as
\be
p_k
&=&\frac{1}{Z}B(H_k)\cr
&=&\frac{1}{Z}\int_0^\infty{\rm d}\beta\,f(\beta)e^{-\beta H_k}
\label {superstatdist}
\ee
with $Z=\sum_kB(H_k)$ (adapting notations to those of the present paper).
The argument in favor of (\ref {superstat}, \ref {superstatdist}) is that in non-equilibrium
systems thermodynamic variables like temperature may fluctuate in space and
time. This argument is not so different from the open systems argument,
given above, that due to interactions with the environment even the
equilibrium distribution will deviate from Boltzmann-Gibbs.

Note that all probability distributions of superstatistics
can be written as distributions of the present generalized thermostatistics.
In other words, any distribution of the form (\ref {superstatdist}))
can be written as (\ref {nonGibbs}) with an appropriate choice
of $\kappa$-deformed exponential function. Indeed, fix some positive temperature $T$
and let
\be
\exp_\kappa(x)=\int_0^\infty{\rm d}\beta\,g(\beta)e^{T\beta x}
\label {ssexp}
\ee
where $g(\beta)$ is a probability distribution still to be determined.
The function $\exp_\kappa(x)$ is increasing and convex.
It satisfies $\exp_\kappa(0)=1$. One has
\be
\int_0^\infty{\rm d}x\,\exp_\kappa(-x)
&=&\int_0^\infty{\rm d}\beta\,g(\beta)\int_0^\infty{\rm d}x\,e^{-T\beta x}\cr
&=&\int_0^\infty{\rm d}\beta\,g(\beta)\frac{1}{T\beta}.
\ee
The latter integral is convergent if the distribution $g(\beta)$ stays away from $\beta=0$.
Then $\exp_\kappa(x)$ satisfies all requirements to be a deformed exponential function in the sense of 
\cite {NJ02}. Now use (\ref {nonGibbs}) to calculate
\be
p_k=\int_0^\infty{\rm d}\beta\,g(\beta)e^{T\beta G(T)-\beta H_k}.
\ee
This expression coincides with (\ref {superstatdist}) provided
\be
g(\beta)=\frac{1}{Z}f(\beta)e^{-\lambda\beta}
\ee
with $\lambda$ the unique solution of
\be
\int_0^\infty{\rm d}\beta\,f(\beta)e^{-\lambda\beta}=Z.
\ee
The normalization function $G(T)$ is then given by
$G(T)=\lambda/T$.


The above arguments show that any probability distribution produced by the
formalism of superstatistics can also be produced by the present formalism
of generalized thermostatistics. Is it possible to distinguish the
two formalisms by applying them to problems of physics? The obvious answer is 
that one should look to the temperature dependence of the equilibrium
probability distributions and compare them with experimental data.
However, superstatistics is meant to be
valid for systems not in equilibrium, with a fluctuating temperature.
It is not at all clear how the distribution of inverse temperatures $f(\beta)$
should depend on some average temperature. This point deserves further
investigation.


The inverse function of $\exp_\kappa(x)$ given by (\ref {ssexp}) is denoted $\ln_\kappa(x)$,
as before. It can be used to calculate an entropy functional $\langle\langle I\rangle\rangle_p$
by means of (\ref {entdef}). Up to a prefactor, this is exactly the entropy functional
introduced in \cite {TS03}, in the first of the two cases considered there.
Indeed, the expression, found there, reads
\be
S(p)=\sum_k\int_0^{p_k}{\rm d}y\,(\alpha+E(y))
\ee
with $\alpha$ a constant and $E(y)$ the inverse of the function $Z^{-1}B(E)$.
In the present notations is
\be
\frac{1}{Z}B(E)=\exp_\kappa\left(\frac{1}{T}(\lambda-E)\right).
\ee
Hence the inverse function is given by
\be
E(y)=\lambda-T\ln_\kappa(y).
\ee
One concludes that
\be
S(p)=\sum_k\int_0^{p_k}{\rm d}y\,(\alpha+\lambda-T\ln_\kappa(y)).
\ee
This equals $T$ times $\langle\langle I\rangle\rangle_p$ provided that
$\alpha$ is chosen equal to $-\lambda-TF_\kappa(0)$.

Note that the formalism developed in \cite {TS03} deviates from the one
presented here because energy constraints other than the standard (\ref{energconstr})
are considered. Such constraints are considered in the final part of this paper.

\subsection {Kaniadakis' entropy functional}

Another example of $\kappa$-deformed logarithm has been introduced
by Kaniadakis \cite{KG01,KG02}, namely
\be
\ln_\kappa(x)=\frac{1}{2\kappa}\left(
x^{\kappa}-x^{-\kappa}
\right),
\label{kaniadakisln}
\ee
with $\kappa$ a parameter with value between -1 and 1.
The corresponding deformed exponential function is
\ben
\exp_\kappa(x)=\left(\kappa x+\sqrt{1+\kappa^2x^2}\right)^{1/\kappa}.
\een
With these definitions the entropy functional (\ref{entdef}) becomes
\ben
\langle\langle I\rangle\rangle_p
&=&
\frac{1}{2\kappa(1-\kappa)}\left(\sum_kp_k^{1-\kappa}-1\right)
+\frac{1}{2\kappa(1+\kappa)}\left(1-\sum_kp_k^{1+\kappa}\right).
\een
The corresponding equilibrium distribution is
\ben
p_k&=&\bigg(\kappa[G(T)-H_k/T]
+\sqrt{1+\kappa^2[G(T)-H_k/T]^2}
\bigg)^{1/\kappa}.
\een
This distribution differs from the one studied in \cite {KG01}
because here normalization is not a prefactor.

\section{Probability-dependent variables}

In this final part of the paper the class of temperature
dependent probabilities (\ref{nonGibbs}) is enlarged.
One cannot expect that all open systems, weakly coupled
with their environment, are described by equilibrium distributions
of the form (\ref{nonGibbs}). Indeed, consider e.g.~an
open system consisting of a finite part of a 2-dimensional Ising
lattice with nearest-neighbor interaction constant $J>0$,
in absence of an external field. Independent of how small $J$
is there is always a critical temperature $T_c$ (proportional to $J$)
where the Ising lattice becomes unstable. This will
be reflected by an instability of the entropy $S(U,B)$
of the open system. The idea is now to capture the essence of
such instabilities by elements coming from mean-field theory.

Temperature-dependent variables are replaced by
probability-dependent variables, which is a new notion
that is introduced here.
By definition, a probability-dependent variable
is a function $f_k(x)$ depending on the microstate $k$ and on the
probability $x$, which varies between 0 and 1. Its average
or expectation is defined by
\ben
\langle\langle f\rangle\rangle_p =\sum_k\int_0^{p_k}{\rm d}x\,f_k(x).
\een
Note that this expression has the essential characteristics
of a statistical average. It is linear, positive on positive
functions, and normalized to one.


The most obvious example of a probability-dependent variable
is information content, also called entropy functional.
Indeed, Shannon's measure of information $I(p)$ can be written as
\ben
I(p)
&=&-\sum_kp_k\ln(p_k)\cr
&=&-\sum_k\int_0^{p_k}{\rm d}x\,(\ln(x)+1)\cr
&=&\langle\langle I\rangle\rangle_p,
\een
with the functions $I_k(x)$ defined by
\ben
I_k(x)=-\ln(x)-1,
\hbox{ independent of }k.
\een
In fact, this is the reason why (\ref{entdef})
was written in the form $\langle\langle I\rangle\rangle_p$, with
the probability-dependent variable $I$ given by
\be
I_k(x)=-\ln_\kappa(x)-F_\kappa(0),
\hbox{ independent of }k.
\label{geninf}
\ee

Notice that also energy may be a probability-dependent variable.
Indeed, consider the following well-known mean-field Hamiltonian
\be
H=-J\sigma \langle\sigma\rangle_p-h\sigma.
\label{meanfieldham}
\ee
In this expression $J>0$ and $h$ are model parameters,
$\sigma$ is a spin variable taking on the value $\pm 1$ with probability
$p_\pm$. The average value of the
Hamiltonian $H$ can be written as
\ben
\langle H\rangle_p
&=&-J\langle\sigma\rangle_p^2-h\langle\sigma\rangle_p\cr
&=&\langle\langle\phi\rangle\rangle_p,
\een
with potentials $\phi_\pm(x)$ given by
\be
\phi_\pm(x)&=&-4Jx+J\mp h.
\label{meanfieldpot}
\ee
This shows that the energy of the mean-field model (\ref{meanfieldham}) is a
probability-dependent variable.

A straightforward generalization of (\ref{meanfieldpot})
is to consider potentials of the form
\be
\phi_k(x)=-J_kqx^{q-1}-h_k,
\label{tsallispot}
\ee
where $q$ is a fixed number, and $J_k>0$ and $h_k$ are model
parameters. The corresponding average energy equals
\ben
\langle\langle\phi\rangle\rangle_p= -\sum_kJ_kp_k^q-\sum_kh_kp_k.
\een
Note that the non-linear constraint (\ref{nonlinenerg})
coincides with this expression, with $J_k<0$ and $h_k=0$.
Hence the critique of \cite {RJD95}, that (\ref{nonlinenerg})
lacks normalization, does not apply. Indeed,
in the present context the average $\langle\langle\phi\rangle\rangle_p$ is
normalized to 1.

\section{Generalized mean-field theory}

The class of temperature-dependent
pdfs (\ref {nonGibbs}) can now be enlarged by allowing all
distributions of the form
\be
p_k=\exp_\kappa(G(T)-\psi_k(p_k)/T),
\label{mfprob}
\ee
where $\psi_k(x)$ is a probability-dependent potential.
The question, discussed below, is whether there exists a generalized mean-field model
with probability-dependent potentials $\phi_k(x)$, for which the equilibrium
distributions are given by (\ref{mfprob}).
The relation between the potentials $\phi_k(x)$
and $\psi_k(x)$ is allowed to be non-trivial.

Let entropy $S$ be defined as before by $S=\langle\langle I\rangle\rangle_p$.
It is immediately clear that internal energy $U$, defined by
$U=\langle\langle \psi\rangle\rangle_p$, still satisfies (\ref{Tdef}). However,
the stability requirement that energy $U$ is an increasing
function of $T$ is not always fulfilled. One can show, e.g., that it holds
when the high temperature condition
\ben
T\ge-\psi'_k(p_k)\exp'_\kappa(G(T)-\psi_k(p_k)/T)
\een
is satisfied for all $k$. It is not a surprise that in these
mean-field models a thermodynamic instability can occur at low
temperatures.


The variational principle is straightforward.
Consider the problem of maximizing $\langle\langle I\rangle\rangle_r$
under the constraint that $\langle\langle\psi\rangle\rangle_r=U$.
This leads to the Lagrange problem of minimizing
\ben
\beta\langle\langle \psi\rangle\rangle_r-\langle\langle I\rangle\rangle_r-\alpha\sum_kr_k.
\een
Variation w.r.t.~$r_k$ leads to the equation
\ben
\beta\psi_k(r_k)-I_k(r_k)=\alpha.
\een
This equation coincides with (\ref{mfprob}), when $r=p$,
$\beta=1/T$, and $\alpha(T)-F_\kappa(0)=G(T)$.


However, the above variational calculation
is {\sl not} what one does traditionally with
the mean-field model (\ref{meanfieldham}).
Rather, one first studies the Hamiltonian $H=-Jm\sigma -h\sigma,$
where $m$ is a constant which is taken equal to
$\langle\sigma\rangle_p$ only after finishing the calculation
of thermodynamic equilibrium. In order to generalize
this procedure, consider the problem of optimizing
$\langle\langle I\rangle\rangle_r$ under the constraint
\ben
\sum_kr_kE_k=U,
\een
with constant energy levels $E_k$. At the end they
are taken equal to $\psi_k(p_k)$.
The corresponding Lagrange problem leads to the equation
\be
\beta E_k-I_k(r_k)=\alpha.
\label{mfeq}
\ee
It is clear that $p_k$ given by (\ref{mfprob}) is
the solution of this problem. However, (\ref{mfeq})
is not the equation corresponding with the
problem of maximizing $\langle\langle I\rangle\rangle_r$
given the constraint $\langle\langle\psi\rangle\rangle_r=U$,
but rather that corresponding with a constraint of the form
$\langle\langle\phi\rangle\rangle_r=U^*$, where the $\phi$-potentials
satisfy
\ben
\psi_k(y)=\frac{1}{y}\int_0^{y}\phi_k(x){\rm d}x.
\een
In the specific case of the mean-field
Hamiltonian (\ref{meanfieldham}), in combination with Shannon's measure
of information, (\ref{mfeq}) reduces to the well known mean-field equation
\ben
m=\tanh\left(\beta(Jm+h)\right)
\qquad\hbox{ with }m=r_+-r_-.
\een


The expression
\be
\beta\sum_kr_kE_k-\langle\langle I\rangle\rangle_r
\label{mffree}
\ee
should be convex, with minimum at $r=p$.
The appropriate tool to investigate this point is relative
information content, also called relative entropy
or Kullback-Leibler distance. Its generalized definition is
(see \cite {NJ03})
\ben
I(r||p)
&=&I(r)-I(p)+\sum_k(r_k-p_k)I_k(p_k).
\een
Using (\ref{geninf}) this becomes
\ben
I(r||p)
&=&\sum_k\int_{p_k}^{r_k}{\rm d}x\,\left(\ln_\kappa(x)-\ln_\kappa(p_k)\right).
\een
A short calculation gives
\ben
\beta\sum_kr_kE_k-\langle\langle I\rangle\rangle_r
=\beta\sum_kp_kE_k-\langle\langle I\rangle\rangle_p
+I(r||p).
\een
Because $I(r||p)$ is a convex function, which is minimal if and only if $r=p$,
this shows that, if a pdf $p$ exists, solving (\ref{mfprob}),
then it minimizes (\ref{mffree}). This does
not mean that the pdf $p$ is the only solution of
(\ref{mfprob}). Indeed, we know that in the mean-field model (\ref{meanfieldham})
there can exist three solutions. But the corresponding energy levels $E_k$ are
different, so that each time (\ref{mffree}) is
minimal at $r=p$.


Let us shortly return to the Tsallis formalism.
Expression (\ref{tsallisprob}) gives
the temperature dependence of the pdf
as found in the context of Tsallis' thermostatistics. It is
straightforward to write these in the form (\ref{mfprob}).
Indeed, let $\psi_k(x)$ be given by
\ben
\psi_k(x)=H_kqx^{q-1}
\een
(note that this is (\ref{tsallispot}) with some change of notation)
and let $\ln_\kappa(x)$ be given by (\ref{tsallisln}).
Then (\ref{mfprob}) reduces to (\ref{tsallisprob}), with
$G(T)$ given by
\ben
G(T)=\frac{q}{q-1}\left(Z(T)^{1-q}-1\right),
\een
and $\beta^*=1/T$.
This means that Tsallis' non-extensive thermostatistics is
a special case of the present generalized mean-field theory.

\section{Discussion}

This paper proposes a generalized thermostatistics obtained by
replacing exponential and logarithmic functions in the
Boltzmann-Gibbs distribution, respectively Shannon's entropy
functional, by $\kappa$-deformed functions. The latter are
rather arbitrary functions satisfying a minimal number of
requirements. In the resulting formalism thermodynamic stability is
automatically satisfied.


New in the present approach is the emphasis on
the temperature dependence of equilibrium probability distributions.
Such temperature dependence may be obtained by
calculation of an open system in the standard formalism,
or may, in principle, originate from an experiment.
Next, a thermodynamic formalism is sought
for in which these are the equilibrium distributions.
This way of working eliminates the need for postulating
specific entropy functions or energy constraints, as is
the typical way of doing in non-extensive thermostatistics.

The thermodynamic relation between entropy $S$ and
internal energy $U$ can be derived from the underlying
microscopic statistical theory under the assumption
that the interaction of the system with its
environment is negligible. Goal of this paper is to show
that this thermodynamic relation is also valid for
systems weakly interacting with their environment.
This has been shown for at least one example,
namely a single spin interacting with its environment.
However, one cannot expect this result to be generic
because the system together with its environment
may exhibit a phase transition. In such a case
one expects that the system might be destabilized
by its environment.


In order to be able to
cope with the latter possibility a mean-field approach has
been elaborated in the second part of the paper.
The concept of probability-dependent variables has been introduced. The result is
a rigorously defined mean-field theory. Its stability properties
have been analyzed using a generalized notion of relative entropy.
Although equilibrium states are not necessarily unique they still
satisfy a variational principle. But the effective energy levels
involved in this variational principle depend on the choice
of equilibrium state.


A secondary goal of the present paper is to improve understanding
of Tsallis' formalism of non-extensive thermostatistics.
With a specific choice of deformed logarithmic function
the entropy functional becomes that of Tsallis. The resulting equilibrium
distribution coincides with Tsallis' distribution.
If temperature dependence of the distributions is taken into account
then the Tsallis formalism describes mean-field models of the
present approach.
As a consequence, all applications of the Tsallis formalism
are also applications of the present generalized thermostatistics.
The same statement holds for the recently introduced formalism
of superstatistics.
Also the entropy functional proposed by Kaniadakis has been considered.
The present formalism leads to an equilibrium distribution
which differs from the one proposed by Kaniadakis because
here normalization is not a prefactor but enters inside the
deformed exponential function.


For simplicity only discrete probabilities have been considered.
Generalization to continuous distributions is straightforward.
The average of a function $f$ over phase space $\Gamma$ is given by
\be
\langle f\rangle_\rho=\int_\Gamma{\rm d}\gamma\,\rho(\gamma)f(\gamma).
\label{contav}
\ee
The analogue of (\ref{nonGibbs}) for the equilibrium density $\rho(\gamma)$ is
\be
\rho(\gamma)=\exp_\kappa\big(G(\beta)-\beta H(\gamma)\big).
\ee
The probability-dependent generalization of (\ref{contav}) is
\be
\langle\langle f\rangle\rangle_\rho=\int_\Gamma{\rm d}\gamma\,\int_0^{\rho(\gamma)}{\rm d}x\,f(\gamma,x).
\ee
Also quantum  statistical mechanics has not been considered.
The {\sl ansatz} for the equilibrium density matrix $\rho$,
given a Hamiltonian operator $H$, is now
\be
\rho=\exp_\kappa\big(G(\beta)-\beta H\big).
\ee
It is not obvious how to introduce probability-dependent variables
in this case. However, for functions of the density matrix, like
information content $I$, one can define
\be
\langle\langle I\rangle\rangle_\rho=\int_0^1{\rm d}x\,\Tr\rho I(x\rho).
\ee
It is then straightforward to reformulate the first part of the paper in a quantum context.

Open systems are most often described using stochastic differential
equations. A recent effort in this direction is found in \cite {CPH03}.
The connection with the present formalism
has not yet been studied.

\section*{Acknowledgments}

I thank S. Abe, G. Kaniadakis, and T. Wada, for
helpful comments on a previous version of this paper.
I thank an anonymous referee for numerous helpful
comments.

\section*{Appendix A: Ising model}

This appendix explains how the deformed logarithm is constructed
in case of the one-dimensional Ising model.

For convenience, let $J^*=J/\Delta$.
Introduce a function $\lambda(x)$, defined for $x$ between 0 and $1/2$,
by the relation
\ben
\lambda\big(p_-(T)\big)=\beta\Delta.
\een
The inverse function $\lambda^{-1}(z)$ is given by
\ben
\lambda^{-1}(z)=\frac{1}{2}\left(
1-\frac{\sinh(z/2)}{\sqrt{e^{-2J^*z}+\sinh^2(z/2)}}\right).
\een
If $J=0$ then an explicit expression for $\lambda(x)$ is feasible.
One finds in this case $\lambda(x)=\ln(1-x)-\ln(x)$.

For $0<x\le 1/2$ the deformed logarithm must be of the form
\be
\ln_\kappa(x)
&=&-\ln\big(2\cosh(\lambda(x)/2)\big)
-u\big(\lambda(x)\big)-\frac{\lambda(x)}{2},\cr
& &
\label{defloglow}
\ee
with $u(z)$ still to be determined. The definition for $1/2\le x<1$
is then given by
\be
\ln_\kappa(x)=\ln_\kappa(1-x)+\lambda(1-x).
\label{defloghigh}
\ee

The derivative of (\ref{defloglow}) reads
\ben
\frac{{\rm d}\,}{{\rm d}x}\ln_\kappa(x)
&=&-\lambda'(x)\bigg[
\frac{1}{2}\left\{1+\tanh\left(\frac{\lambda(x)}{2}\right)\right\}
+u'(\lambda(x))
\bigg].
\een
Because $\lambda(x)$ is decreasing and the deformed logarithm must be an increasing function
one obtains the condition that
\ben
2u'(z)+\tanh(z/2)> -1.
\een
Because $\ln_\kappa(x)$ must also increase for $1/2\le x <1$ one has further that
\ben
2u'(z)+\tanh(z/2)<1.
\een

The condition that $\ln_\kappa(1)=0$ can be analyzed easily. The result is that
$u(z)$ must vanish in the limit of large $z$. The condition that
$\ln_\kappa(x)$ is concave in $x=1/2$ requires that $u'(0)\ge 0$.

The condition of concavity of $\ln_\kappa(x)$ is more difficult
because $\lambda(x)$ is necessarily convex for $x$ close to zero,
while $\lambda''(1/2)=-16J^*$.
This wrong curvature of $\lambda(x)$ close to $x=1/2$ should
be compensated by the curvature of $u(z)$ for $z$ close to zero.

The above analysis suggests to choose $u$ as follows
\be
u(z)=\frac{1}{2}J^*(1+z)e^{-z}.
\label{uchoice}
\ee
The factor $J^*$ has been included to ensure that the deformed
logarithm coincides with the natural logarithm in case $J^*=0$.
The above conditions on $u(z)$ are satisfied provided $J^*$ is less than
$J^*_{\rm max}\simeq 3.83$.

It is rather hard to check concavity of the deformed logarithm
in an analytic manner. However, by numerical evaluation
one can convince oneself that the function is concave as long as
$J^*$ is not too large.

\section*{Appendix B: Thermodynamic stability}

In this appendix is shown that the probability distributions
considered in the present paper are automatically stable
under thermodynamic perturbations.
It is a tradition (see \cite {CHB85}) to require also that
pressure is a decreasing function of volume. However,
here the notion of volume dependence is absent.

Consider the problem of maximizing entropy $\langle\langle I\rangle\rangle_r$
under the constraints that average energy $\langle H\rangle$
and average magnetization $\langle M\rangle$ have given values $U$, respectively $B$:
\ben
\langle H\rangle&=&\sum_k r_k H_k=U\cr
\langle M\rangle&=&\sum_k r_k M_k=B.
\label{appbconstr}
\een
Introduce Lagrange multipliers $\alpha$, $\beta$, and $\gamma$,
and minimize the expression
\ben
\beta\sum_k r_kH_k+\gamma\sum_k r_kM_k-\langle\langle I\rangle\rangle_r-\alpha\sum_kr_k.
\een
Variation w.r.t.~$r_k$ gives the condition
\ben
0=\beta H_k+\gamma M_k+\ln_\kappa(r_k)+F_\kappa(0)-\alpha.
\een
Hence the probability distribution maximizing $\langle\langle I\rangle\rangle_r$
must be of the form
\be
\ln_\kappa(r_k)=-F_\kappa(0)+\alpha -\beta H_k-\gamma M_k.
\label{appbeqd}
\ee
The parameter $\alpha$ must be such that $\sum_kr_k=1$
and will be considered as a function of $\beta$ and $\gamma$.
The value of $\langle\langle I\rangle\rangle_r$ at equilibrium is denoted $S(\beta,\gamma)$.

Introduce escort probabilities $P_k(\beta,\gamma)$ by
\ben
P_k(\beta,\gamma)
&=&\frac{1}{Z(\beta,\gamma)}
\exp'_\kappa(-F_\kappa(0)+\alpha -\beta H_k-\gamma M_k).\cr
& &
\een
where
\ben
Z(\beta,\gamma)=\sum_k\exp'_\kappa(-F_\kappa(0)+\alpha -\beta H_k-\gamma M_k).
\een
Then, using (\ref{appbeqd}), one obtains
\ben
\frac{\partial U}{\partial \beta}
&=&Z(\beta,\gamma)\sum_kP_kH_k\left[\frac{\partial \alpha}{\partial \beta}-H_k\right]\cr
\frac{\partial U}{\partial \gamma}
&=&Z(\beta,\gamma)\sum_kP_kH_k\left[\frac{\partial \alpha}{\partial \gamma}-M_k\right]\cr
\frac{\partial B}{\partial \beta}
&=&Z(\beta,\gamma)\sum_kP_kM_k\left[\frac{\partial \alpha}{\partial \beta}-H_k\right]\cr
\frac{\partial B}{\partial \gamma}
&=&Z(\beta,\gamma)\sum_kP_kM_k\left[\frac{\partial \alpha}{\partial \gamma}-M_k\right].
\een
Now, from $\sum_kr_k=1$ follows
\ben
\frac{\partial \alpha}{\partial \beta}&=&\langle H\rangle_*\cr
\frac{\partial \alpha}{\partial \gamma}&=&\langle M\rangle_*,
\een
where we introduced the notation that $\langle X\rangle_*=\sum_kP_kX_k$.

This leads to the result that
\ben
\frac{\partial U}{\partial \beta}
&=&-Z(\beta,\gamma)\left[\langle H^2\rangle_*-\langle H\rangle_*^2\right]\cr
\frac{\partial U}{\partial \gamma}
&=&\frac{\partial B}{\partial \beta}
=-Z(\beta,\gamma)\left[\langle HM\rangle_*-\langle H\rangle_*\,\langle M\rangle_*\right]\cr
\frac{\partial B}{\partial \gamma}
&=&-Z(\beta,\gamma)\left[\langle M^2\rangle_*-\langle M\rangle_*^2\right].
\een

A short calculation using (\ref{appbeqd}) and (\ref{entdef}) gives
\be
\frac{\partial S}{\partial \beta}
&=&\beta\frac{\partial U}{\partial \beta}+\gamma\frac{\partial B}{\partial \beta}\cr
\frac{\partial S}{\partial \gamma}
&=&\beta\frac{\partial U}{\partial \gamma}+\gamma\frac{\partial B}{\partial \gamma}.
\label{appbderivs}
\ee
Introduce the notation
\ben
D(\beta,\gamma)
&=&\frac{\partial U}{\partial \beta}\frac{\partial B}{\partial\gamma}
-\frac{\partial U}{\partial \gamma}\frac{\partial B}{\partial\beta}.
\een
Then one has, using (\ref{appbderivs}),
\ben
\frac{\partial S}{\partial U}
&=&\frac{1}{D(\beta,\gamma)}
\left[
\frac{\partial S}{\partial \beta}\frac{\partial B}{\partial\gamma}
-\frac{\partial S}{\partial \gamma}\frac{\partial B}{\partial\beta}\right]
\cr
&=&\beta
\een
and similarly
\ben
\frac{\partial S}{\partial B}
&=&\gamma.
\een

\vskip 2cm
Let us now put everything together. The matrix of second derivatives
of $S$ as a function of $U$ and $B$ equals
\ben
& &\left(
\begin{array}{lr}
\frac{\partial^2 S}{\partial U^2} &\frac{\partial^2 S}{\partial B\partial U}\\
\frac{\partial^2 S}{\partial U\partial B} & \frac{\partial^2 S}{\partial B^2}\\
\end{array}
\right)
=
\frac{1}{D(\beta,\gamma)}
\left(
\begin{array}{lr}
\frac{\partial B}{\partial\gamma} &-\frac{\partial U}{\partial \gamma}\\
-\frac{\partial B}{\partial\beta} &\frac{\partial U}{\partial \beta}\\
\end{array}
\right)\cr
&=&-\frac{Z(\beta,\gamma)}{D(\beta,\gamma)}
\left(
\begin{array}{lr}
\langle M^2\rangle_*-\langle M\rangle_*^2
&-\langle HM\rangle_*+\langle H\rangle_*\,\langle M\rangle_*\\
-\langle HM\rangle_*+\langle H\rangle_*\,\langle M\rangle_*
&\langle H^2\rangle_*-\langle H\rangle_*^2 \\
\end{array}
\right).
\label{covmat}
\ee
From Schwartz's inequality follows that
\ben
|-\langle HM\rangle_*+\langle H\rangle_*\,\langle M\rangle_*|^2
\le
\left[\langle M^2\rangle_*-\langle M\rangle_*^2
\right]\,\left[
\langle H^2\rangle_*-\langle H\rangle_*^2
\right].
\een
This inequality suffices to show that the matrix in the r.h.s.~of (\ref{covmat})
has positive eigenvalues and that $D(\beta,\gamma)\ge 0$.
Hence, entropy is a concave function of $U$ and $B$. This shows
thermodynamic stability of the probability distribution $r$.



\begin{thebibliography}{99}
\raggedright

\bibitem {TC88} C. Tsallis, {\sl Possible Generalization of
Boltzmann-Gibbs Statistics,}
J. Stat. Phys. {\bf 52}, 479-487 (1988).

\bibitem {BC03} C. Beck, E.G.D. Cohen, {\sl Superstatistics},
Physica A322, 267-275 (2003).

\bibitem {ST99} {\sl Non-extensive statistical mechanics and thermodynamics,} eds.
S.R.A. Salinas and C. Tsallis, Braz. J. Phys. {\bf 29}(1) (1999).

\bibitem{TC94} C. Tsallis,
{\sl What are the numbers that experiments provide?}
Quimica Nova {\bf 17}, 468 (1994)

\bibitem {KG01} G. Kaniadakis,
{\sl Nonlinear kinetics underlying generalized statistics,}
Physica A{\bf 296}, 405-425 (2001).

\bibitem{KA30}A. Kolmogorov, Atti della R. Accademia Nazionale dei
Lincei {\bf 12}, 388 (1930).

\bibitem{NM30}M. Nagumo, Japan. J. Math. {\bf 7}, 71 (1930).

\bibitem {LPS94a} J.T. Lewis, C.-E. Pfister, W.G. Sullivan,
{\sl Large deviations and the thermodynamic formalism:
a new proof of the equivalence of ensembles,} in:
{\sl On three levels,} ed. M. Fannes, C. Maes, A. Verbeure
(Plenum press, New York, 1994)

\bibitem {LPS94b} J.T. Lewis, C.-E. Pfister, W.G. Sullivan,
{\sl The equivalence of ensembles for lattice systems:
some examples and a counterexample,}
J. Stat. Phys. {\bf 77}(1/2), 397-419 (1994).

\bibitem {LP95} J.T. Lewis and C.-E. Pfister,
{\sl Thermodynamic probability theory: some
aspects of large deviations,}
Russian Math. Surveys, {\bf 50}(2), 279-317 (1995).

\bibitem {GR77} R.M. Gray, {\sl Entropy and information theory}
(Springer Verlag, 1990)

\bibitem {NJ02} J. Naudts, {\sl Deformed exponentials and logarithms
in generalized thermostatistics,} arXiv:cond-mat/0203489,
Physica A{\bf 316}, 1-12 (2002).

\bibitem {KW41} H.A. Kramers, G.H. Wannier, {\sl Statistics of the twodimensional
ferromagnet. Part I,} Phys. Rev. {\bf 60}, 252-276 (1941).

\bibitem {NJ03} J. Naudts, {\sl Continuity of $\kappa$-deformed
entropies and relative entropies,} arXiv:math-ph/0208038v2.

\bibitem {BS93} C. Beck, F. Schl\"ogl, {\sl Thermodynamics of chaotic systems:
An introduction} (Cambridge University Press, Cambridge, 1993)

\bibitem {TMP98} C. Tsallis, R.S. Mendes, A.R. Plastino,
{\sl The role of constraints within generalized non-extensive
statistics,} Physica A{\bf 261}, 543-554 (1998).

\bibitem {CHB85} H. B. Callen, {\sl Thermodynamics and an introduction
to thermostatistics} (John Wiley \& Sons, 1985),
Chapter 8.

\bibitem {AMPP01} S. Abe, A. Martinez, F. Pennini, A. Plastino,
{\sl Nonextensive thermodynamics relations,}
Phys. Lett. A{\bf 281}(2-3), 126-130 (2001).

\bibitem {CT91} E.M.F. Curado, C. Tsallis,
{\sl Generalized statistical mechanics: connection with thermodynamics,}
J. Phys. A {\bf24}, L69-72 (1991).

\bibitem {RJD95} J.D. Ramshaw, {\sl Thermodynamic stability conditions for the
Tsallis and R\'enyi entropies,} Phys. Lett. A{\bf 198}, 119-121 (1995).

\bibitem {TC95} C. Tsallis, {\sl Comment on ``Thermodynamic stability conditions for the
Tsallis and R\'enyi entropies'' by J.D. Ramshaw,}
Phys. Lett. A{\bf 206}, 389-391 (1995).

\bibitem {WT02} T. Wada, {\sl On the thermodynamic stability conditions
of Tsallis' entropy,} Phys. Lett. A{\bf 297}, 334-337 (2002).

\bibitem {WW00} G. Wilk, Z. Wlodarczyk, {\sl Interpretation of the non-extensivity
parameter q in some applications of Tsallis statistics and L\'evy distributions,}
Phys. Rev. Lett. {\bf 84}, 2770-2773 (2000).

\bibitem {BC01} C. Beck, {\sl Dynamical foundations of nonextensive statistical mechanics,}
Phys. Rev. Lett. {\bf 87}, 180601 (2001).

\bibitem {Be03} C. Beck, {\sl Superstatistics: Theory and Applications,} arXiv:cond-mat/0303288.

\bibitem {TS03} C. Tsallis, A.M.C. Souza, {\sl Constructing a statistical mechanics
for Beck-Cohen superstatistics,} Phys. Rev. E{\bf 67}, 026106 (2003).

\bibitem {KG02} G. Kaniadakis, {\sl Statistical mechanics in the
context of special relativity,} Phys. Rev. E{\bf 66}, 056125 (2002).

\bibitem {CPH03} P.-H. Chavanis, {\sl Generalized thermodynamics and
kinetic equations: Boltzmann, Landau, Kramers and Smoluchowski,}
arXiv:cond-mat/0304073v1.

\end{thebibliography}
\end{document}